\documentstyle[pra,aps,floats,epsfig]{revtex}
\begin{document}
\draft
\title{Possibility of the tunneling time determination}
\author{Julius Ruseckas}
\address{Institute of Theoretical Physics and Astronomy,\\
A. Go\v{s}tauto 12, 2600 Vilnius, Lithuania}
\date{\today}
\maketitle

\begin{abstract}
We show that it is impossible to determine the time a tunneling particle
spends under the barrier. However, it is possible to determine the
asymptotic time, i.e., the time the particle spends in a large area
including the barrier. We propose a model of time measurements. The model
provides a procedure for calculation of the asymptotic tunneling and
reflection times. The model also demonstrates the impossibility of
determination of the time the tunneling particle spends under the barrier.
Examples for $\delta$-form and rectangular barrier illustrate the obtained
results.
\end{abstract}

\section{Introduction}

Tunneling phenomena are inherent in numerous quantum systems, ranging from atom
and condensed matter to quantum fields. Therefore, the questions about the
tunneling mechanisms are important. There have been many attempts to define
a physical time for tunneling processes, since this question has been raised by
MacColl \cite{maccol} in 1932. This question is still the subject of much
controversy, since numerous theories contradict each other in their
predictions for ``the tunneling time''. Some of these theories predict that
the tunneling process is faster than light, whereas the others state that it
should be subluminal. This subject has been covered in a number of reviews
(Hauge and St\o vneng \cite{haugestovneng}, 1989;
Olkholovsky and Recami \cite{olkhovskyrecami}, 1992; Landauer and Martin
\cite{landauermartin}, 1994 and Chiao and Steinberg \cite{chiao}, 1997). The
fact that there is a time related to the tunneling process has been observed
experimentally \cite
{gueret1,gueret2,esteve,enders,ranfagni,spielmann,heitmann,balcou,garrison}.
However, the results of the experiments are ambiguous.

Many of the theoretical approaches can be divided into three categories.
First, one can study evolution of the wave packets through the barrier and get
the phase time. However, the correctness of the definition of this time is
highly questionable \cite{buttikerlandauer}. Another approach is based on
the determination of a set of dynamic paths, i.e., the calculation of the time
the different paths spend in the barrier and averaging over the set of the
paths. The paths can be found from the Feynman path integral formalism
\cite{sokolovskibaskin}, from the Bohm approach \cite
{leavens1,leavens2,leavens3,leavens4}, or from the Wigner distribution \cite
{muga}. The third class uses a physical clock which can be used for
determination of the time elapsed during the tunneling (B\"{u}ttiker and
Landauer used an oscillatory barrier \cite{buttikerlandauer},
Baz' suggested  the Larmor time \cite{baz}).

The problems rise also from the fact, that the arrival time of a particle
to the definite spatial point is a classical concept. Its quantum counterpart
is problematic, even for the free particle case. In classical mechanics for
the determination of the time the particle spends moving along a certain
trajectory we have to measure the position of the particle at two different
moments of time. In quantum mechanics this procedure does not work. From
Heisenberg's uncertainty principle it follows that we cannot measure the
position of a particle without alteration of its momentum. To determine
exactly the arrival time of a particle, one has to measure the
position of the particle with a great precision. Because of the measurement,
the momentum of the particle will have a big uncertainty and the second
measurement will be indefinite. If we want to ask about the time in quantum
mechanics, we need to define the procedure of the measurement. We can measure
the position of the particle only with a finite precision and get a
distribution of the possible positions. Applying such a measurement, we can
expect to obtain not a single value of the traversal time but a distribution of
times.

The question {\it How much time does the tunneling particle spend in the
barrier region?} is not precise. There are two different, but related questions
connected with the tunneling time problem \cite{dumontmarchioro}:

\begin{enumerate}
\item[(i)]  How much time does the tunneling particle spend under the
barrier?
\item[(ii)]  At what time does the particle arrive to the point behind the
barrier?
\end{enumerate}

There have been many attempts to answer these questions. However, there are
several papers showing, that according to quantum mechanics the question (i)
makes no sense \cite {dumontmarchioro,sokolovskiconnor,yamada,BSM}. The goal of
this paper is to investigate the possibility to determine the tunneling time
using a concrete model of time measurements.

The paper is organized as follows: In section \ref{secimpos} we prove that it
is impossible to determine the time the tunneling particle spends under the
barrier. In section \ref{secmod} we present the procedure of the time
measurement. This procedure leads to the dwell time if no distinctions between
tunneled and reflected particles are made. This is shown in section
\ref{secdwell}. In section \ref{seccondprob} we modify the proposed procedure
of the time measurement to make the distinction between tunneled and reflected
particles and obtain the tunneling time. The result of such a procedure clearly
shows the impossibility of the determination of the tunneling time. However, it
also gives the method of the asymptotic time calculation. In sections
\ref{secprop} and \ref{secrefltime} we examine the properties of the tunneling
and reflection times. In section \ref{secasympt}, we derive the formula for
asymptotic time. Section \ref{secconcl} summarizes our findings.

\section{Impossibility of the tunneling time determination}
\label{secimpos}

To answer the question {\it How much time does the tunneling particle spend
under the barrier?} we need a criterion of the tunneling. In this paper we
accept the following criterion: the particle had tunneled in the case it was in
front of the barrier at first and later it was found behind the barrier. We require
that the mean energy of the particle and the energy uncertainty must be less
than the height of the barrier. Following this criterion, we introduce an
operator corresponding to the ``tunneling-flag'' observable. This operator
projects the wave function onto the subspace of functions localized behind the
barrier
\begin{equation}
\hat{f}_{T}\left( X\right) =\Theta \left( \hat{x}-X\right)  \label{tunflag}
\end{equation}
where $\Theta $ is the Heaviside unit step function and $X$ is a point
behind the barrier. We call operator $\hat{f}_{T}$ as the tunneling flag
operator. This operator has two eigenvalues: $0$ and $1$. The eigenvalue $0$
corresponds to the fact that the particle has not tunneled, while the
eigenvalue $1$ corresponds to the tunneled particle.

We will work in the Heisenberg's representation. In this representation, the
tunneling flag operator is
\begin{equation}
\tilde{f}_{T}\left( t,X\right) =\exp \left( \frac{i}{\hbar }\hat{H}t\right)
\hat{f}_{T}\left( X\right) \exp \left( -\frac{i}{\hbar }\hat{H}t\right) .
\label{tunflagheis}
\end{equation}
To take into account all tunneled particles, the limit $t\rightarrow
+\infty $ must be taken. So, the ``tunneling-flag'' observable in the
Heisenberg's picture is represented by the operator $\tilde{f}_{T}\left(
\infty ,X\right)=\lim_{t\rightarrow +\infty}\tilde{f}_{T}\left(t,X\right)$.
We can obtain explicit expression for this operator.

The operator $\tilde{f}_{T}\left( t,X\right) $ obeys the equation
\begin{equation}
i\hbar \frac{\partial }{\partial \ t}\tilde{f}_{T}\left( t,X\right) =\left[
\tilde{f}_{T}\left( t,X\right) ,\hat{H}\right] . \label{eq:ft}
\end{equation}
The commutator in Eq.\ (\ref{eq:ft}) may be expressed as
\[
\left[ \tilde{f}_{T}\left( t,X\right) ,\hat{H}\right] =\exp \left( \frac{i}{
\hbar }\hat{H}t\right) \left[ \hat{f}_{T}\left( X\right) ,\hat{H}\right]
\exp \left( -\frac{i}{\hbar }\hat{H}t\right) .
\]
If the Hamiltonian has the form $\hat{H}=\frac{1}{2M}\hat{p}^{2}+V\left(
\hat{x}\right) $ then the commutator takes the form
\begin{equation}
\left[ \hat{f}_{T}\left( X\right) ,\hat{H} \right] =i\hbar \hat{J}\left(
X\right)
\end{equation}
where $\hat{J}\left( X\right) $ is the probability flux operator
\begin{equation}
\hat{J}\left( x\right) =\frac{1}{2M}\left( \left| x\right\rangle
\left\langle x\right| \hat{p}+\hat{p}\left| x\right\rangle \left\langle
x\right| \right) .  \label{probflux}
\end{equation}
Therefore, we have equation for the commutator
\begin{equation}
\left[\tilde{f}_{T}\left(t,X\right),\hat{H}\right]
=i\hbar\tilde{J}\left(X,t\right) . \label{eq:com}
\end{equation}
The initial condition for the function $\tilde{f\left(t,X\right)}$ may be
defined as
\[
\tilde{f}_{T}\left(t=0,X\right)=\hat{f}_{T}\left(X\right) .
\]
From Eqs.\ (\ref{eq:ft}) and\ (\ref{eq:com}) we obtain equation for the
evolution of the tunneling flag operator
\begin{equation}
i\hbar \frac{\partial }{\partial \ t}\tilde{f}_{T}\left( t,X\right) =i\hbar
\tilde{J}\left( X,t\right) . \label{eq:evol}
\end{equation}
From Eq.\ (\ref{eq:evol}) and initial condition it follows an explicit
expression for the tunneling flag operator
\begin{equation}
\tilde{f}_{T}\left( t,X\right) =\hat{f}_{T}\left( X\right)
+\int_{0}^{t}dt_{1}\tilde{J}\left( X,t_{1}\right) .  \label{tunflagflux}
\end{equation}

In the question ``How much time does the tunneling particle spend under the
barrier?'' we ask about the particles, which we know with certainty
have tunneled. In addition, we want to have some information about the
location of the particle. However, does quantum mechanics allow us to have
the information about the tunneling and location simultaneously? A
projection operator
\begin{equation}
\hat{D}\left( \Gamma \right) =\int\limits_{\ \Gamma }dx\left| x\right\rangle
\left\langle x\right|  \label{posop}
\end{equation}
where $\left| x\right\rangle $ is the eigenfunction of the coordinate
operator, represents the probability for the particle to be in the region
$\Gamma$. In the Heisenberg's representation this operator takes the form
\begin{equation}
\tilde{D}\left(\Gamma ,t\right)=\exp\left(\frac{i}{\hbar}\hat{H}t\right)\hat{D}
\left(\Gamma\right)\exp\left(-\frac{i}{\hbar}\hat{H}t\right). \label{posopheis}
\end{equation}
From Eqs.\ (\ref{probflux}),\ (\ref{tunflagflux}) and\ (\ref{posopheis}) we see
that the operators $\tilde{D}\left(\Gamma ,t\right)$ and $\tilde{f}_{T}
\left(\infty ,X\right)$ in general do not commute. This means that we cannot
simultaneously have the information about the tunneling and location of the
particle. If we know with certainty that the particle has tunneled then we can
say nothing about its location in the past and if we know something about the
location of the particle, we cannot determine definitely whether the particle
will tunnel. Therefore, the question {\it How much time does the tunneling
particle spend under the barrier}? cannot be answered in principle, if the
question is so posed that its precise definition requires the existence of the
joint probability that the particle is found in $\Gamma$ at time $t$ and
whether or not it is found on the right side of the barrier at a sufficiently
later time. A similar analysis has been performed in Ref. \cite{BSM}. It has
been shown that due to non-commutability of operator there exist no unique
decomposition of the dwell time.

This conclusion is, however, not only negative. We know that $\int_{-\infty}^{+
\infty}dx\left|x\right\rangle\left\langle x\right|=1$ and $\left[1,\tilde{f}_{T}
\left(\infty ,X\right)\right]=0$. Therefore, if the region $\Gamma $ is large
enough, one has a possibility to answer the question about the
tunneling time.

From the fact that the operators $\tilde{D}\left(\Gamma ,t\right)$ and
$\tilde{f}_{T}\left(\infty ,X\right)$ do not commute we can predict that the
measurement of the tunneling time will yield a value dependent on the
particular detection scheme. The detector is made so that it yields some value.
But if we try to measure non-commuting observables, the measured values depends
on the interaction between the detector and the measured system. So, in the
definition of the Larmor time there is a dependence on the type of boundary
attributed to the magnetic-field region \cite{olkhovskyrecami}.

\section{The model of the time measurement}
\label{secmod}

We consider a model for the tunneling time measurement which is somewhat
similar to the ``gedanken'' experiment used to obtain the Larmor time, but
it is simpler and more transparent. This model had been proposed by
A. M. Steinberg \cite{steinberg}, however, it was treated in non-standard
way, introducing complex probabilities. Here we use only the formalism of
the standard quantum mechanics.

\begin{figure}[htpb]
\begin{center}
\epsfxsize=.5\hsize \epsffile{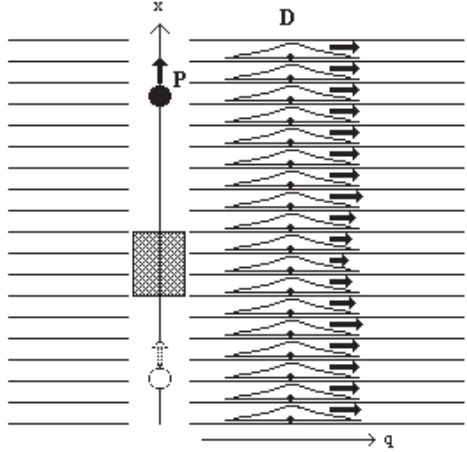}
\end{center}
\caption{The configuration of the measurements of the tunneling time. The
particle {\bf P} is tunneling along $x$ coordinate and it is interacting with
detectors {\bf D}. The barrier is represented by the rectangle. The interaction
with the definite detector occurs only in the narrow region limited by the
horizontal lines. The changes of the momenta of the detectors are represented
by arrows.}
\label{config}
\end{figure}
Our system consists of particle {\bf P} and several detectors {\bf D}. Each
detector interacts with the particle only in the narrow region of space. The
configuration of the system is shown in Fig.\ \ref{config}. When the
interaction of the particle with the detectors is weak the detectors do not
influence the state of the particle. Therefore, we can analyze the action of
detectors separately.

First of all we consider the interaction of the particle with one detector. The
Hamiltonian of the system is
\begin{equation}
\hat{H}=\hat{H}_{P}+\hat{H}_{D}+\hat{H}_{I}
\end{equation}
where $\hat{H}_{P}=\frac{1}{2M}\hat{p}^{2}+V\left(\hat{x}\right)$ is the
Hamiltonian of the particle, $\hat{H}_{D}$ is the detector's Hamiltonian and
\begin{equation}
\hat{H}_{I}=\gamma \hat{q}\hat{D}\left( x_{D}\right)  \label{interact}
\end{equation}
represents the interaction between the particle and the detector. The operator
$\hat{q}$ acts in the Hilbert space of the detector. We require a continuous
spectrum of the operator $\hat{q}$. For simplicity, we can consider this
operator as the coordinate of the detector. The operator $\hat{D}\left(
x_{D}\right) $ acts in the Hilbert space of the particle. In the coordinate
representation it is nonvanishing only in the small region around the point
$x_{D}$. In an ideal case the operator $\hat{D}\left( x_{D}\right) $ may be
expressed as $\delta$-function of the particle coordinate,
\begin{equation}
\hat{D}\left( x_{D}\right) \equiv \left| x_{D}\right\rangle \left\langle
x_{D}\right| =\delta \left( \hat{x}-x_{D}\right) .  \label{delta}
\end{equation}
Parameter $\gamma $ in Eq.\ (\ref{interact}) characterizes the strength of
the interaction. A very small parameter $\gamma $ ensures the undisturbance
of the particle's motion.

Hamiltonian\ (\ref{interact}) with $\hat{D}$ given by\ (\ref{delta}) represents
the constant force acting on the detector {\bf D} when the particle is very
close to the point $x_{D}$. This force results in the change of the detector's
momentum. From the classical point of view, the change of the momentum is
proportional to the time the particle spends in the region around $x_{D}$ and
the coefficient of proportionality equals to the force acting on the detector.
In the ordinary quantum mechanics there is no general method of the time
determination. If we want to define such a method, we have to make additional
assumptions about the time. It is natural to extend the classical method of the
time determination into the quantum mechanics, too. Therefore we assume that
the change of the mean momentum of the detector is proportional to the time the
constant force acts on the detector and that the time the particle spends in
the detector's region is the same as the time the force acts on the detector.

We can replace the $\delta$-function by the narrow rectangle of the width
$L$ and height $\frac{1}{L}$. From\ (\ref{interact}) it follows that the
force acting on the detector when the particle is in the region around $x_{D}$
is $-\gamma\frac{1}{L}$. The time the particle spends in the region
around $x_{D}$ equals to $\left(-\gamma\frac{1}{L}\right)^{-1}\left(
\left\langle p_{q}\left(t\right)\right\rangle -\left\langle
p_{q}\right\rangle\right)$ where $p_{q}$ is the momentum of the detector
conjugated to the coordinate $q$, while $\left\langle p_q\right\rangle$ and
$\left\langle p_q\left(t\right)\right\rangle$ are the mean initial momentum and
momentum after time $t$, respectively. The time the particle spends until
time moment $t$ in the unit length region is
\begin{equation}
\tau \left( t \right) = -\frac{1}{\gamma}\left( \left\langle p_{q}
\left( t\right) \right\rangle -\left\langle p_{q}\right\rangle \right) .
\label{timedef}
\end{equation}
To find the time the particle spends in the region of the finite length, we
have to add the times spent in the regions of length $L$. When
$L\rightarrow 0$ we obtain an integral.

The evolution operator is
\begin{equation}
\hat{U}\left( t\right) =\exp \left[ -\frac{i}{\hbar }\left( \hat{H}_{P}+\hat{
H}_{D}+\hat{H}_{I}\right) t\right] .
\end{equation}
In the moment $t=0$ the density matrix of the whole system is $\hat{\rho}
\left(0\right)=\hat{\rho}_{P}\left(0\right)\otimes\hat{\rho}_{D}\left(
0\right)$ where $\hat{\rho}_{P}\left(0\right)$ is the density matrix of
the particle and $\hat{\rho}_{D}\left(0\right)=\left|\Phi\right\rangle
\left\langle\Phi\right|$ is the density matrix of the detector with $\left|
\Phi\right\rangle$ being the normalized vector in the Hilbert space of the
detector. After the interaction, the density matrix of the detector is $\hat{
\rho}_{D}\left( t\right) =Tr_{P}\left\{ \hat{U}\left( t\right) \left( \hat{
\rho}_{P}\left( 0\right) \otimes \left| \Phi \right\rangle \left\langle \Phi
\right| \right) \hat{U}^{+}\left( t\right) \right\} $. In the moment $t=0$ it
must be $\left\langle\mathstrut x\right|\rho_{P}\left(\mathstrut 0\right)
\left|x^{\prime}\right\rangle\neq 0$ only when $x<0$ and $x^{\prime}<0$.

Further, for simplicity we will neglect the Hamiltonian of the detector. The
evolution operator then approximately equals to the operator $\hat{U}\left(
t,\gamma\hat{q}\right)$ where
\begin{equation}
\hat{U}\left(t,\alpha\right)=\exp\left[-\frac{i}{\hbar}\left(\frac{1}{2M}
\hat{p}^{2}+V\left(\hat{x}\right)+\alpha\hat{D}\left(x_{D}\right)
\right) t\right] .  \label{aproxevol}
\end{equation}
After such assumptions from our model we can obtain the time, the particle
spends in the definite space region. Similar calculations were done for
detector's position rather than momentum by G. Iannaccone \cite{iann}.

\section{Measurement of the dwell time}
\label{secdwell}

We expand the operator $\hat{U}\left(t,\gamma\hat{q}\right)$ into the
series of the parameter $\gamma$, assuming that $\gamma $ is small. Introducing
the operator $\hat{D}\left(x_{D}\right)$ in the interaction representation
\begin{equation}
\tilde{D}\left(x_{D},t\right)=\exp\left(\frac{i}{\hbar}\hat{H}_{P}
t\right)\hat{D}\left(x_{D}\right)\exp\left(-\frac{i}{\hbar}
\hat{H}_{P}t\right)
\end{equation}
we obtain the first order approximation for the operator $\hat{U}\left(
t,\gamma\hat{q}\right)$,
\begin{equation}
\hat{U}\left(t,\gamma\hat{q}\right)\approx\exp\left(-\frac{i}{\hbar}
\hat{H}_{P}t\right)\left(1+\frac{\gamma\hat{q}}{i\hbar}
\int_{0}^{t}dt_{1}\tilde{D}\left(x_{D},t_{1}\right)\right) .
\end{equation}
For shortening the notation we introduce an operator
\begin{equation}
\hat{F}\left(x_{D},t\right)\equiv\int_{0}^{t}dt_{1}\tilde{D}\left(
x_{D},t_{1}\right)  \label{opf}
\end{equation}
and the equation for the evolution operator $\hat{U}\left(t,\gamma\hat{q}
\right)$ is expressed as
\begin{equation}
\hat{U}\left(t,\gamma\hat{q}\right)\approx\exp\left(-\frac{i}{\hbar}
\hat{H}_{P}t\right)\left(1+\frac{\gamma\hat{q}}{i\hbar}\hat{F}\left(
x_{D},t\right)\right) .  \label{evolop1}
\end{equation}
The density matrix of the detector in the coordinate representation in the
first order approximation then is
\begin{eqnarray*}
\left\langle q\right|\rho_{D}\left(t\right)\left|q^{\prime}\right
\rangle &=& \langle q\left|\Phi\right\rangle\langle\Phi\left|
q^{\prime}\right\rangle Tr\left\{\hat{U}\left(t,\gamma q\right)
\hat{\rho}_{P}\left(0\right)\hat{U}^{\dag}\left(t,\gamma q^{\prime}\right)
\right\} \\
&=& \langle q\left|\Phi\right\rangle\langle\Phi\left|q^{\prime}
\right\rangle\left(1+\frac{\gamma q}{i\hbar}\left\langle\hat{F}\left(
x_{D},t\right)\right\rangle-\frac{\gamma q^{\prime}}{i\hbar}\left\langle
\hat{F}\left(x_{D},t\right)\right\rangle\right) \\
& \approx & \left\langle q\right|\exp\left[-\frac{i}{\hbar}\gamma
\left\langle\hat{F}\left(x_{D},t\right)\right\rangle\hat{q}\right]
\left|\Phi\right\rangle\left\langle\Phi\right|\exp\left[\frac{i}{\hbar}
\gamma\left\langle\hat{F}\left(x_{D},t\right)\right\rangle\hat{q}
\right]\left|q^{\prime}\right\rangle .
\end{eqnarray*}
The average momentum of the detector after time $t$ is $\left\langle
p_{q}\right\rangle-\gamma\left\langle\hat{F}\left(x_{D},t\right)
\right\rangle$ where $\left\langle\ p_{q}\right\rangle=\left\langle\Phi
\right|\hat{p}_{q}\left|\Phi\right\rangle$ and $\left\langle\hat{F}\left(
x_{D},t\right)\right\rangle=Tr\left\{\hat{F}\left(x_{D},t\right)
\hat{\rho}_{P}\left(0\right)\right\}$. From Eq.\ (\ref{timedef}) we obtain
the time the particle spends in the unit length region between time
momentum $t=0$ and $t$
\begin{equation}
\tau ^{\text{Dw}}\left( x,t\right) =\left\langle \hat{F}\left( x,t\right)
\right\rangle .  \label{dwelltime}
\end{equation}
The time spent in the space region restricted by the coordinates $x_{1}$ and
$x_{2}$ is
\begin{equation}
t^{\text{Dw}}\left( x_{2},x_{1}\right) = \int_{x_{1}}^{x_{2}}dx\,\tau
^{\text{Dw}}\left( x,t\rightarrow \infty \right) = \int_{x_{1}}^{x_{2}}
dx\int_{0}^{\infty }\rho \left( x,t\right) dt.
\end{equation}
This is a well-known expression for the dwell time \cite{olkhovskyrecami}.
The dwell time is the average over entire ensemble of particles regardless they
are tunneled or not. The expression for the dwell time obtained in our model
is the same as the well-known expression obtained by other authors. Therefore,
we can expect that our model can yield a reasonable expression for the
tunneling time as well.

\section{Conditional probabilities and the tunneling time}
\label{seccondprob}

Having seen that our model gives the time averaged over the entire ensemble of
particles, let us now take the average only over the subensemble of the
tunneled particles. The joint probability that the particle has tunneled {\it
and} the detector has the momentum $p_{q}$ at the time moment $t$ is
$W\left(f_{T}, p_q;t\right)=Tr\left\{\hat{f}_T
\left(X\right)\left|p_q\right\rangle\left\langle
p_q\right|\hat{\rho}\left(t\right)\right\}$ where $\left|p_{q} \right\rangle$
is the eigenfunction of the momentum operator $\hat{p}_{q}$ and the tunneling
flag operator $\hat{f}_{T}\left(X\right)$ is defined by Eq.\ (\ref{tunflag}).
In quantum mechanics such a probability does not always exist. If the joint
probability does not exist then the concept of the conditional probability is
meaningless. But in our case the operators $\hat{f}_{T}\left(X\right)$ and
$\left|p_{q}\right\rangle \left\langle p_{q}\right|$ commute, therefore, the
probability $W\left(f_{T}, p_{q};t\right)$ exists. The conditional probability
that the momentum of the detector is $p_{q}$ provided that the particle has
tunneled is given according to the Bayes's theorem, i.e.,
\begin{equation}
W\left(p_{q};t\left|f_{T}\right.\right)=\frac{W\left(f_{T},p_{q};
t\right)}{W\left(f_{T};t\right)}  \label{condprob}
\end{equation}
where $W\left(f_{T};t\right)=Tr\left\{\hat{f}_{T}\left(X\right)\hat{\rho}
\left(t\right)\right\}$ is the probability that the particle has tunneled
until time $t$. The average momentum of the detector with the
condition that the particle has tunneled is
\[
\left\langle p_{q}\left(t\right)\right\rangle =\int p_{q}dp_{q}W\left(
p_{q};t\left|f_{T}\right.\right)
\]
or
\begin{equation}
\left\langle p_{q}\left( t\right) \right\rangle =\frac{1}{W\left( f_{T};\
t\right) }Tr\left\{ \hat{f}_{T}\left( X\right) \hat{p}_{q}\hat{\rho}\left(
t\right) \right\} .  \label{condave}
\end{equation}

In the first order approximation the probability $W\left( f_{T};t\right) $
is given by equation
\begin{equation}
W\left( f_{T};t\right) \approx \left\langle \tilde{f}_{T}\left( t,X\right)
\right\rangle + \frac{\gamma }{i\hbar }\left\langle q\right\rangle
\left\langle \left[ \tilde{f}_{T}\left( t,X\right) ,\hat{F}\left(
x_{D},t\right) \right] \right\rangle .  \label{tunprob}
\end{equation}
The expression $Tr\left\{\hat{f}_{T}\left(X\right)\hat{p}_{q}\hat{\rho}
\left(t\right)\right\}$ in Eq.\ (\ref{condave})in the first order approximation
reads
\begin{eqnarray*}
Tr\left\{ \hat{f}_{T}\left( X\right) \hat{p}_{q}\hat{\rho}\left(
t\right) \right\} & \approx &  \left\langle p_{q}\right\rangle
\left\langle \tilde{f}_{T}\left( t,X\right) \right\rangle \\
&&+ \frac{\gamma }{i\hbar }\left( \left\langle \tilde{f}_{T}\left( t,X\right)
\hat{F}\left(x_{D},t\right) \right\rangle \left\langle \hat{p}_{q}\hat{q}
\right\rangle - \left\langle \hat{q}\hat{p}_{q}\right\rangle \left\langle
\hat{F}\left(x_{D},t\right)\tilde{f}_{T}\left( t,X\right) \right\rangle \right) .
\end{eqnarray*}
Using the commutator $\left[ \hat{q},\hat{p}_{q}\right] =i\hbar $ from Eqs. \
(\ref{timedef}) and\ (\ref{condave}) we obtain the time the tunneled particle
spends in the unit length region around $x$ until time $t$
\begin{eqnarray}
\tau \left(x,t\right) &=& \frac{1}{2\left\langle \tilde{f}_{T}\left( t,X\right)
\right\rangle } \left\langle \tilde{f}_{T}\left( t,X\right) \hat{F}
\left( x,t\right) +\hat{F} \left( x,t\right) \tilde{f}_{T}\left( t,X\right)
\right\rangle  \nonumber \\
&& +\frac{1}{i\hbar \left\langle \tilde{f}_{T}\left( t,X\right)
\right\rangle }\left( \left\langle q\right\rangle \left\langle
p_{q}\right\rangle-\text{Re}\left\langle\hat{q}\hat{p}_{q}\right\rangle \right)
\left\langle \left[ \tilde{f}_{T}\left( t,X\right) ,\hat{F}\left( x,t\right)
\right] \right\rangle .  \label{tuntime}
\end{eqnarray}
The obtained expression\ (\ref{tuntime}) for the tunneling time is real, on the
contrary to the complex-time approach. It should be noted that this expression
even in the limit of the very weak measurement depends on the particular
detector. This yields from the impossibility of the determination of the
tunneling time. If the commutator $\left[\tilde{f}_{T}\left(t,X\right),\hat{F}
\left(x,t\right)\right]$ is zero, the time has a precise value. If the
commutator is not zero, only the integral of this expression over a large
region has the meaning of an asymptotic time related to the large region as we
will see in Sec. \ref{secasympt}.

Eq.\ (\ref{tuntime}) can be rewritten as a sum of two terms, the first
term being independent of the detector and the second dependent, i.e.,
\begin{equation}
\tau\left( x,t\right)=\tau^{\text{Tun}}\left(x,t\right)+\frac{2}{\hbar}\left(
\left\langle q\right\rangle\left\langle p_{q}\right\rangle-\text{Re}
\left\langle\hat{q}\hat{p}_{q}\right\rangle\right)\tau_{\text{corr}}^{
\text{Tun}}\left(x,t\right)
\end{equation}
where
\begin{mathletters}
\label{tuntimeparts}
\begin{eqnarray}
\tau^{\text{Tun}}\left(x,t\right)  &=&\frac{1}{2\left\langle
\tilde{f}_{T}\left(t,X\right)\right\rangle}\left\langle\tilde{f}_{T}\left(
t,X\right)\hat{F}\left(x,t\right)+\hat{F}\left(x,t\right)
\tilde{f}_{T}\left(t,X\right)\right\rangle ,  \label{tunteimre} \\
\tau_{\text{corr}}^{\text{Tun}}\left(x,t\right)  &=&\frac{1}{
2\,i\left\langle\tilde{f}_{T}\left(t,X\right)\right\rangle}\left\langle
\left[\tilde{f}_{T}\left(t,X\right) ,\hat{F}\left(x,t\right)\right]
\right\rangle .  \label{tuntimeim}
\end{eqnarray}
\end{mathletters}
The quantities $\tau^{\text{Tun}}\left(x,t\right)$ and
$\tau_{\text{corr}}^{\text{Tun}}\left(x,t\right)$ do not depend on the
detector.

In order to separate the tunneled and reflected particles we have to take the
limit $t\rightarrow \infty $. Otherwise, the particles which tunnel after the
time $t$ would not contribute to the calculation. So we introduce operators
\begin{mathletters}
\label{opinf}
\begin{eqnarray}
\hat{F}\left( x\right)  &=&\int_{0}^{\infty }dt_{1}\tilde{D}\left(
x,t_{1}\right) ,  \label{opefinf} \\
\hat{N}\left( x\right)  &=&\int_{0}^{\infty }dt_{1}\tilde{J}\left(
x,t_{1}\right) .  \label{opN}
\end{eqnarray}
\end{mathletters}
From Eq.\ (\ref{tunflagflux}) it follows that the operator
$\tilde{f}_{T}\left(\infty ,X\right)$ is equal to $\hat{f}_{T}\left(X\right)
+\hat{N}\left( X\right)$. As long as the particle is initially before the
barrier
\[
\hat{f}_{T}\left( X\right) \hat{\rho}_{P}\left( 0\right) =
\hat{\rho}_{P}\left( 0\right) \hat{f}_{T}\left( X\right) =0.
\]
In the limit $t\rightarrow \infty $ we have
\begin{mathletters}
\label{tuntimeinf}
\begin{eqnarray}
\tau^{\text{Tun}}\left(x\right) &=&\frac{1}{2\left\langle\hat{N}\left(X\right)
\right\rangle}\left\langle\hat{N}\left(X\right)\hat{F}\left(x\right)+\hat{F}
\left(x\right)\hat{N}\left(X\right)\right\rangle , \label{tuntimereinf} \\
\tau_{\text{corr}}^{\text{Tun}}\left(x\right) &=&\frac{1}{2\,i\left\langle
\hat{N}\left(X\right)\right\rangle}\left\langle\left[\hat{N}\left(X\right)
,\hat{F}\left(x\right)\right]\right\rangle .  \label{tuntimeiminf}
\end{eqnarray}
\end{mathletters}

Let us define an ``asymptotic time'' as the integral of $\tau(x,\infty)$ over a
wide region containing the barrier. Since the integral of
$\tau_{\text{corr}}^{\text{Tun}}\left(x\right)$ is very small compared to that
of $\tau^{\text{Tun}}\left(x\right)$ as we will see later, the asymptotic time
is effectively the integral of $\tau^{\text{Tun}}\left(x\right)$ only. This
allows us to identify $\tau ^{\text{Tun}}\left( x\right)$ as ``the density of
the tunneling time''.

In many cases for the simplification of mathematics it is common to write the
integrals over time as the integrals from $-\infty$ to $+\infty$. In our model
we cannot without additional assumptions integrate in Eqs.\ (\ref{opinf}) from
$-\infty$, because the negative time means the motion of the particle to the
initial position. If some particle in the initial wave packet had negative
momentum then in the limit $t\rightarrow -\infty $ it was behind the barrier
and contributed to the tunneling time.

\section{Properties of the tunneling time}
\label{secprop}

As it has been mentioned above, the question {\it How much time does a
tunneling particle spend under the barrier} has no exact answer. We can
determine only the time the tunneling particle spends in a large region,
containing the barrier. In our model this time is expressed as an integral of
quantity\ (\ref{tuntimereinf}) over the region. In order to determine the
properties of this integral it is useful to determine properties of the
integrand.

To be able to expand the range of integration over time to $-\infty$,
it is necessary to have the initial wave packet far to the left from the
points under the investigation and this wave packet must consist only of the
waves moving in the positive direction.

It is convenient to make calculations in the energy representation.
Eigenfunctions of the Hamiltonian $\hat{H}_P$ are $\left|E,\alpha\right\rangle$,
where $\alpha =\pm 1$. The sign '$+$' or '$-$' corresponds to
the positive or negative initial direction of the wave, respectively. Outside
the barrier these eigenfunctions are:
\begin{mathletters}
\label{eigenfunct}
\begin{eqnarray}
\langle x\left| E,+\right\rangle &=&\left\{
\begin{array}{l}
\sqrt{\frac{M}{2\pi \hbar p_{E}}}\left( \exp \left( \frac{i}{\hbar }
p_{E}x\right) +r\left( E\right) \exp \left( -\frac{i}{\hbar }p_{E}x\right)
\right) \text{, }x<0, \\
\sqrt{\frac{M}{2\pi \hbar p_{E}}}t\left( E\right) \exp \left( \frac{i}{\hbar
}p_{E}x\right) \text{, }x>L,
\end{array}
\right .  \label{eigenplus} \\
\langle x\left| E,-\right\rangle &=&\left\{
\begin{array}{l}
\sqrt{\frac{M}{2\pi \hbar p_{E}}}t\left( E\right) \exp \left( -\frac{i}{
\hbar }p_{E}x\right) \text{, }x<0, \\
\sqrt{\frac{M}{2\pi \hbar p_{E}}}\left( \exp \left( -\frac{i}{\hbar }
p_{E}x\right) -\frac{t\left( E\right) }{t^{*}\left( E\right) }r^{*}\left(
E\right) \exp \left( \frac{i}{\hbar }p_{E}x\right) \right) \text{, }x>L
\end{array}
\right .  \label{eigenminus}
\end{eqnarray}
\end{mathletters}
where $t\left( E\right) $ and $r\left( E\right) $ are transmission and
reflection amplitudes respectively,
\begin{equation}
p_{E}=\sqrt{2ME},  \label{impuls}
\end{equation}
the barrier is in the region between $x=0$ and $x=L$ and $M$ is the mass of
the particle. These eigenfunctions are orthonormal, i.e.,
\begin{equation}
\langle E,\alpha\left|E^{\prime},\alpha^{\prime}\right\rangle =\delta
_{\alpha,\alpha^{\prime}}\delta\left(E-E^{\prime}\right) .
\label{orthogonality}
\end{equation}
The evolution operator is
\[
\hat{U}_{P}\left( t\right) =\sum_{\alpha }\int_{0}^{\infty }dE\left|
E,\alpha \right\rangle \left\langle E,\alpha \right| \exp \left( -\frac{i}{
\hbar }Et\right) .
\]
The operator $\hat{F}\left( x\right) $ is given by the equation
\[
\hat{F}\left( x\right) =\int_{-\infty }^{\infty }dt_{1}\sum_{\alpha ,\alpha
^{\prime }} \int\!\!\!\int dE\,dE^{\prime }\left| E,\alpha \right\rangle
\left\langle E,\alpha \right| x\rangle \langle x\left| E^{\prime },
\alpha ^{\prime}\right\rangle \left\langle E^{\prime },\alpha ^{\prime }\right|
\exp \left( \frac{i}{\hbar }\left( E-E^{\prime }\right) t_{1}\right)
\]
where the integral over the time is $2\pi\hbar\delta\left(E-E^{\prime}\right)$
and, therefore,
\[
\hat{F}\left( x\right) =2\pi \hbar \sum_{\alpha ,\alpha ^{\prime }}\int
dE\left| E,\alpha \right\rangle \left\langle E,\alpha \right| x\rangle
\langle x\left| E,\alpha ^{\prime }\right\rangle \left\langle E,\alpha
^{\prime }\right| .
\]
In analogous way
\[
\hat{N}\left( x\right) =2\pi \hbar \sum_{\alpha ,\alpha
^{\prime }}\int dE\left| E,\alpha \right\rangle \left\langle E,\alpha
\right| \hat{J}\left( x\right) \left| E,\alpha ^{\prime }\right\rangle
\left\langle E,\alpha ^{\prime }\right| .
\]
We consider the initial wave packet consisting only of the waves moving in
the positive direction. Then we have
\begin{eqnarray*}
\left\langle \hat{N}\left( x\right) \right\rangle  &=&2\pi \hbar \int
dE\left\langle \left| E,+\right\rangle \left\langle E,+\right| \hat{J}\left(
x\right) \left| E,+\right\rangle \left\langle E,+\right| \right\rangle , \\
\left\langle \hat{F}\left( x\right) \hat{N}\left( X\right) \right\rangle
&=&4\pi ^{2}\hbar ^{2}\sum_{\alpha }\int dE\left\langle \left|
E,+\right\rangle \left\langle E,+\right| x\rangle \langle x\left| E,\alpha
\right\rangle \left\langle E,\alpha \right| \hat{J}\left( X\right) \left|
E,+\right\rangle \left\langle E,+\right| \right\rangle .
\end{eqnarray*}
From the condition $X>L$ it follows
\begin{equation}
\left\langle \hat{N}\left( X\right) \right\rangle =\int dE\left\langle
\left| E,+\right\rangle \left| t\left( E\right) \right| ^{2}\left\langle
E,+\right| \right\rangle .  \label{Nave}
\end{equation}
For $x<0$ we obtain the following expressions for the quantities
$\tau^{\text{Tun}}\left(x,t\right)$ and $\tau_{\text{corr}}^{\text{Tun}}\left(
x,t\right)$
\begin{mathletters}
\label{tuntimebefore}
\begin{eqnarray}
\tau^{\text{Tun}}\left(x,t\right) &=&\frac{M}{\left\langle \hat{N}\left(
X\right) \right\rangle }\int dE\left\langle \left| E,+\right\rangle \frac{1}{
2p_{E}}\left| t\left( E\right) \right| ^{2}\left( 2+r\left( E\right) \exp
\left( -2\frac{i}{\hbar }p_{E}x\right) \right. \right. \nonumber \\
&& + \left. \left. r^{*}\left( E\right) \exp \left( 2%
\frac{i}{\hbar}p_{E}x\right)\right)\left\langle E,+\right|\right\rangle , \\
\tau_{\text{corr}}^{\text{Tun}}\left(x,t\right) &=&\frac{M}{2\left\langle
\hat{N}\left( X\right) \right\rangle }\int dE\left\langle \left|
E,+\right\rangle \frac{1}{ip_{E}}\left| t\left( E\right) \right| ^{2}\left(
r\left( E\right) \exp \left( -2\frac{i}{\hbar }
p_{E}x\right) \right. \right. \nonumber \\
&& - \left. \left. r^{*}\left( E\right) \exp
\left( 2\frac{i}{\hbar }p_{E}x\right) \right) \left\langle E,+\right|
\right\rangle .
\end{eqnarray}
\end{mathletters}
For $x>L$ these expressions take the form
\begin{mathletters}
\label{tuntimeafter}
\begin{eqnarray}
\tau ^{\text{Tun}}\left( x,t\right)  &=&\frac{M}{\left\langle \hat{N}\left(
X\right) \right\rangle }\int dE\left\langle \left| E,+\right\rangle \frac{1}{
2p_{E}}\left| t\left( E\right) \right| ^{2}\left( 2-\frac{t\left( E\right) }{
t^{*}\left( E\right) }r^{*}\left( E\right) \exp \left( 2\frac{i}{\hbar }
p_{E}x\right) \right. \right. \nonumber \\
&& - \left. \left. \frac{t^{*}\left( E\right) }{t\left( E\right) }r\left(
E\right) \exp \left( -2\frac{i}{\hbar }p_{E}x\right) \right) \left\langle
E,+\right| \right\rangle , \\
\tau _{\text{corr}}^{\text{Tun}}\left( x,t\right)  &=&\frac{M}{2\left\langle
\hat{N}\left( X\right) \right\rangle }\int dE\left\langle \left| E,+
\right\rangle \frac{i}{p_{E}}\left| t\left( E\right) \right| ^{2}\left(
\frac{t\left( E\right) }{t^{*}\left( E\right) }r^{*}\left( E\right) \exp \left(
2\frac{i}{\hbar }p_{E}x\right) \right. \right. \nonumber \\
&& - \left. \left. \frac{t^{*}\left( E\right) }{t\left( E\right) }r\left(
E\right) \exp \left( -2\frac{i}{\hbar }p_{E}x\right) \right) \left\langle
E,+\right| \right\rangle .
\end{eqnarray}
\end{mathletters}

We illustrate the obtained formulae for the $\delta$-function barrier
\[
V\left( x\right) =\Omega \delta \left( x\right)
\]
and for the rectangular barrier. The incident wave packet is Gaussian and it is
localized far to the left from the barrier.

\begin{figure}[htbp]
\begin{center}
\epsfxsize=.5\hsize \epsffile{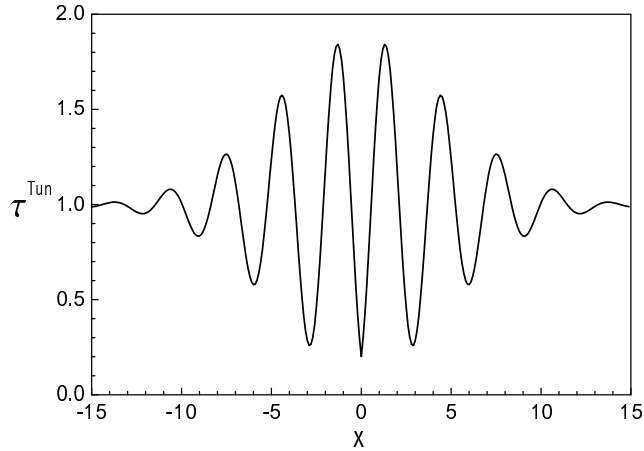}
\end{center}
\caption{``The asymptotic time density'' for $\delta$-function barrier with the
parameter $\Omega=2$. The barrier is located at the point $x=0$. Units are such
that $\hbar =1$ and $M=1$ and the average momentum of the Gaussian wave packet
$\left\langle p\right\rangle=1$. In these units length and time are
dimensionless. The width of the wave packet in the momentum space $\sigma
=0.001$.}
\label{tundelta}
\end{figure}
\begin{figure}[htbp]
\begin{center}
\epsfxsize=.5\hsize \epsffile{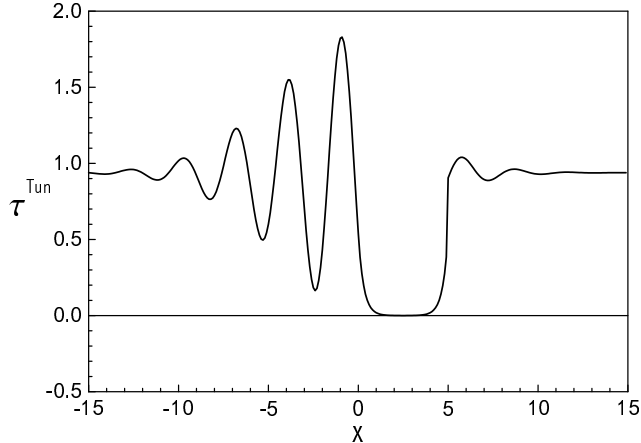}
\end{center}
\caption{``The asymptotic time density'' for rectangular barrier. The barrier
is localized between the points $x=0$ and $x=5$ and the height of the barrier
is $V_{0}=2$. The used units and parameters of the initial wave packet are the
same as in Fig.\ \protect\ref{tundelta}.}
\label{tunrect}
\end{figure}
In Fig. \ref{tundelta} and \ref{tunrect}, we see interference-like
oscillations near the barrier. Oscillations are not only in front of the
barrier but also behind the barrier. When $x$ is far from the barrier the
``time density'' tends to the value close to $1$. This is in agreement with
classical mechanics because in the chosen units the mean velocity of the
particle is $1$. In Fig \ref{tunrect}, another property of ``tunneling time
density'' is seen: it is almost zero in the barrier region. This explains the
Hartmann and Fletcher effect \cite{hartmann,fletcher}: for opaque barriers the
effective tunneling velocity is very large.

\section{The reflection time}
\label{secrefltime}

We can easily adapt our model for the reflection, too. For doing this, we
should replace the tunneling-flag operator $\hat{f}_{T}$ by the reflection
flag operator
\begin{equation}
\hat{f}_{R}=1-\hat{f}_{T}.  \label{reflflag}
\end{equation}
Replacing $\hat{f}_{T}$ by $\hat{f}_{R}$ in Eqs.\ (\ref{tuntimeinf}) we
obtain the equality
\begin{equation}
\left\langle \tilde{f}_{R}\left( t=\infty ,X\right) \right\rangle \tau
^{\text{Refl}}\left( x\right) =\tau ^{\text{Dw}}\left( x\right) -\left\langle
\tilde{f}_{T}\left( t=\infty ,X\right) \right\rangle \tau ^{\text{Tun}}\left(
x\right) .
\label{importcond}
\end{equation}
We see that in our model the important condition
\begin{equation}
\tau^{\text{Dw}}=T \tau^{\text{Tun}}+R \tau^{\text{Refl}}
\end{equation}
where $T$ and $R$ are transmission and reflection probabilities is
satisfied automatically.

If the wave packet consist only of the waves moving in positive direction, the
density of dwell time is
\begin{equation}
\tau ^{\text{Dw}}\left( x,t\right) =2\pi \hbar \int dE\left\langle \left|
E,+\right\rangle \left\langle E,+\right| x\rangle \langle x\left|
E,+\right\rangle \left\langle E,+\right| \right\rangle .
\label{dwelltimeaprox}
\end{equation}
For $x<0$ we have
\begin{eqnarray}
\tau ^{\text{Dw}}\left( x,t\right) &=& M\int dE\left\langle \left| E,
+\right\rangle \frac{1}{p_{E}}\left( 1+\left| r\left( E\right)
\right| ^{2}+r\left( E\right) \exp \left( -2\frac{i}{\hbar }p_{E}x\right)
\right. \right. \nonumber \\
&& + \left. \left. r^{*}\left( E\right)
\exp \left( 2\frac{i}{\hbar }p_{E}x\right) \right) \left\langle E,+\right|
\right\rangle
\end{eqnarray}
and for the reflection time we obtain the ``time density''
\begin{eqnarray}
\tau ^{\text{Refl}}\left( x\right) &=& \frac{M}{1-\left\langle \hat{N}\left(
X\right)\right\rangle }\int dE\left\langle \left| E,+\right\rangle
\frac{1}{p_{E}}\left( 2\left| r\left( E\right) \right| ^{2} \right. \right.
\nonumber \\
&& + \left. \left. \frac{1}{2}\left( 1+\left|r\left( E\right) \right| ^{2}
\right) r\left(E\right) \exp \left(-2\frac{i}{\hbar }p_{E}x\right) +
r^{*}\left( E\right) \exp \left( 2\frac{i}{\hbar }%
p_{E}x\right) \right) \left\langle E,+\right| \right\rangle .
\end{eqnarray}
For $x>L$ the density of the dwell time is
\begin{equation}
\tau ^{\text{Dw}}\left( x,t\right) =M\int dE\left\langle \left| E,
+\right\rangle \frac{1}{p_{E}}\left| t\left( E\right) \right| ^{2}\left\langle
E,+\right| \right\rangle
\end{equation}
and the ``density of the reflection time'' may be expressed as
\begin{eqnarray}
\tau ^{\text{Refl}}\left( x\right) &=& \frac{M}{2}\int dE\left\langle \left|
E,+\right\rangle \frac{1}{p_{E}}\left| t\left( E\right) \right| ^{2}
\left(\frac{t\left( E\right) }{t^{*}\left( E\right) }r^{*}\left( E\right) \exp
\left( 2\frac{i}{\hbar }p_{E}x\right) \right. \right. \nonumber \\
&& + \left. \left. \frac{t^{*}\left( E\right) }{t\left(
E\right) }r\left( E\right) \exp \left( -2\frac{i}{\hbar }p_{E}x\right)
\right) \left\langle E,+\right| \right\rangle .
\end{eqnarray}

We illustrate the properties of the reflection time for the same barriers. The
incident wave packet is Gaussian and it is localized far to the left from the
barrier.

\begin{figure}[htbp]
\begin{center}
\epsfxsize=.5\hsize \epsffile{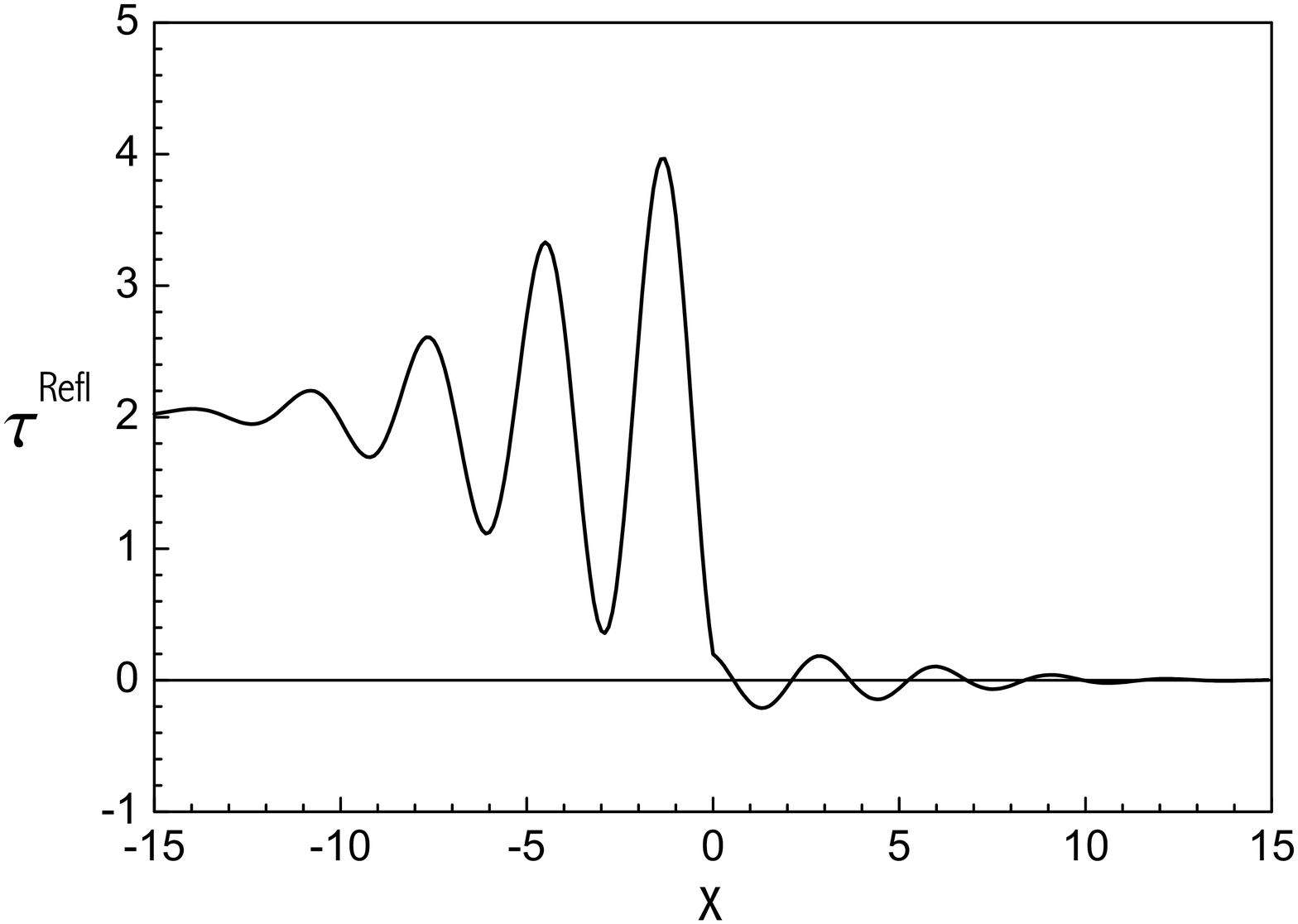}
\end{center}
\caption{``Reflection time density'' for the same conditions as in Fig.\
\protect\ref{tundelta}.}
\label{refldelt}
\end{figure}
\begin{figure}[htbp]
\begin{center}
\epsfxsize=.5\hsize \epsffile{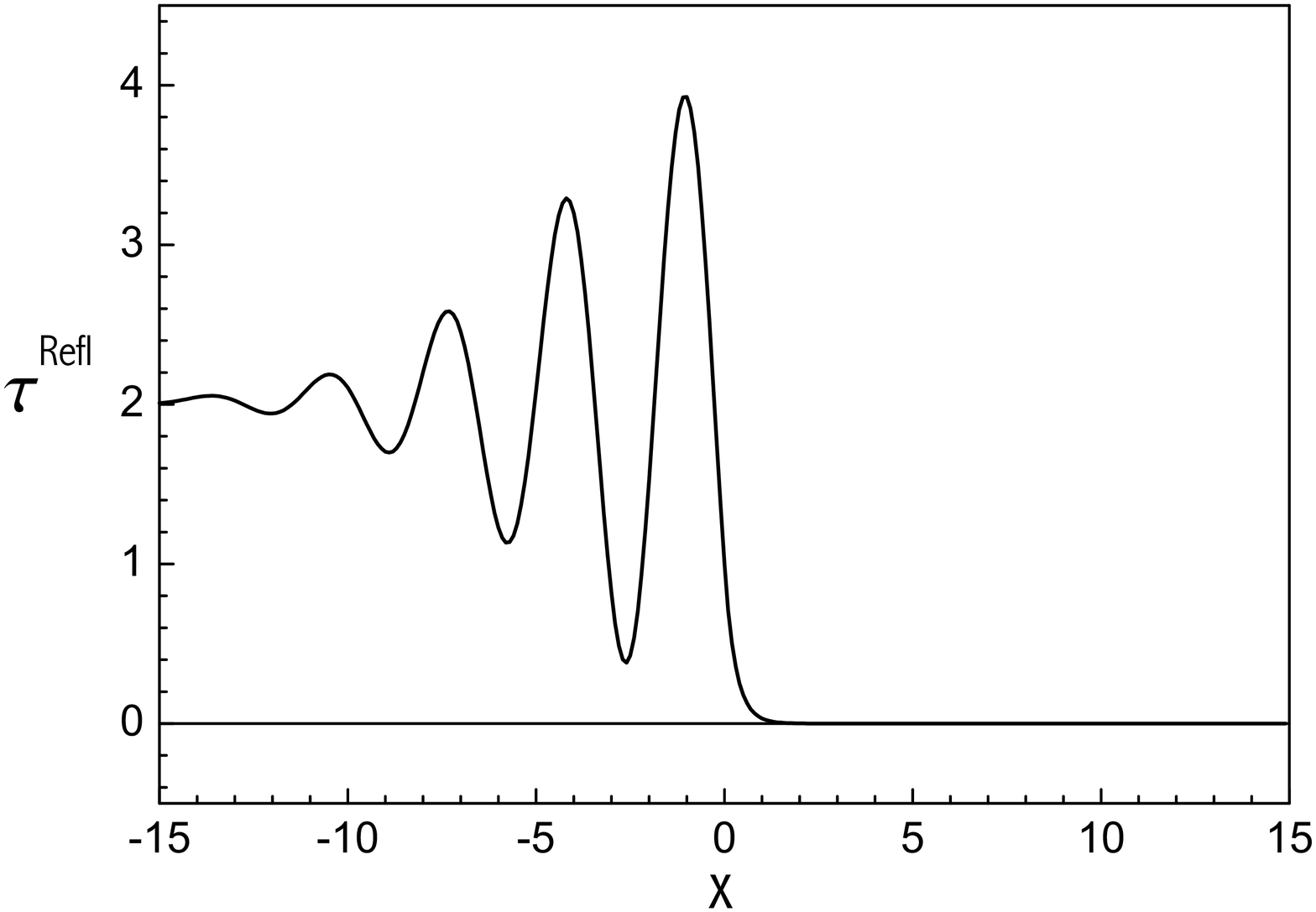}
\end{center}
\caption{``Reflection time density'' for the same conditions as in Fig.\
\protect\ref{tunrect}.}
\label{reflrect}
\end{figure}
\begin{figure}[htbp]
\begin{center}
\epsfxsize=.5\hsize \epsffile{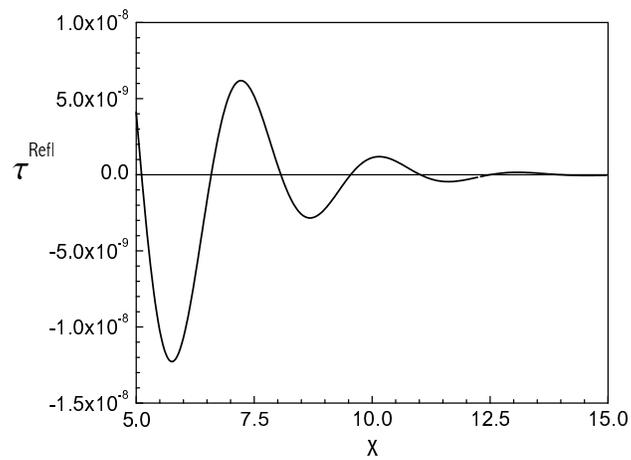}
\end{center}
\caption{``Reflection time density'' for rectangular barrier in the area behind
the barrier. The parameters and the initial conditions are the same as in Fig.\
\protect\ref{tunrect} }
\label{reflrect2}
\end{figure}
In Fig. \ref{refldelt} and \ref{reflrect}, we also see the interference-like
oscillations at both sides of the barrier. As far as for the rectangular
barrier the ``time density'' is very small, the part behind the barrier is
presented in Fig. \ref{reflrect2}. Behind the barrier, the ``time density'' in
certain places becomes negative. This is because the quantity $\tau
^{\text{Refl}}\left( x\right) $ is not positive definite. Non-positivity is the
direct consequence of non-commutativity of operators in Eqs.\ (\ref
{tuntimeinf}). There is nothing strange in negativity of $\tau
^{\text{Refl}}\left( x\right) $, because this quantity itself has no physical
meaning. Only the integral over the large region has the meaning of time. When
$x$ is far to the left from the barrier the ``time density'' tends to the value
close to $2$ and when $x$ is far to the right from the barrier the ``time
density'' tends to $0$. This is in agreement with classical mechanics because
in the chosen units the velocity of the particle is $1$ and the reflected
particle crosses the area before the barrier two times.

\section{The asymptotic time}
\label{secasympt}

As it was mentioned above, we can determine only the time which the
tunneling particle spends in a large region containing the barrier, i.e.,
the asymptotic time. In our model this time is expressed as an integral of
quantity\ (\ref{tuntimereinf}) over this region. We can do the integration
explicitly.

The continuity equation yields
\begin{equation}
\frac{\partial}{\partial t}\tilde{D}\left(x_D,t\right)+\frac{\partial
}{\partial x_D}\tilde{J}\left( x_{D},t\right) =0.
\end{equation}
The integration in Eq.\ (\ref{opf}) can  be performed by parts
\[
\int_{0}^{t}dt_{1}\tilde{D}\left(x_{D},t_{1}\right)=t\tilde{D}\left(
x_{D},t\right) +\frac{\partial }{\partial x}\int_{0}^{t}t_{1}dt_{1}\tilde{J}
\left( x_{D},t_{1}\right) .
\]
If the density matrix $\hat{\rho}_{P}\left(
0\right) $ represents localized particle then $\lim_{t\rightarrow \infty
}\left( \tilde{D}\left( x,t\right) \hat{\rho}_{P}\left( 0\right) \right) =0$.
The operator $\tilde{D}\left( x,t\right) $ in all expressions under
consideration is multiplied by $\hat{\rho}_{P}\left( 0\right) $. Therefore
we can write an effective equality
\begin{equation}
\int_0^{\infty}dt_1\tilde{D}\left(x_{D},t_1\right)=\frac{\partial}{\partial
x}\int_0^{\infty}t_1 dt_1 \tilde{J}\left(x_{D},t_1 \right).
\end{equation}
We introduce the operator
\begin{equation}
\hat{T}\left( x\right) =\int_{0}^{\infty }t_{1}dt_{1}\tilde{J}\left(
x,t_{1}\right) .  \label{opT}
\end{equation}
We consider the asymptotic time, i.e., the time the particle spends between
points $x_{1}$ and $x_{2}$ when $x_{1}\rightarrow -\infty $, $x_{2}
\rightarrow +\infty $,
\[
t^{\text{Tun}}\left(x_{2},x_{1}\right)=\int_{x_{1}}^{x_{2}}dx\,\tau
^{\text{Tun}}\left(x\right).
\]
After the integration we have
\begin{equation}
t^{\text{Tun}}\left( x_{2},x_{1}\right) =t^{\text{Tun}}\left( x_{2}\right)
-t^{\text{Tun}}\left(x_{1}\right)   \label{tuntimeasympt}
\end{equation}
where
\begin{equation}
t^{\text{Tun}}\left( x\right) =\frac{1}{2\left\langle \hat{N}\left( x\right)
\right\rangle }\left\langle \hat{N}\left( x\right) \hat{T}\left( x\right) +
\hat{T}\left( x\right) \hat{N}\left( x\right) \right\rangle .
\label{asympttimeterm}
\end{equation}

If we assume that the initial wave packet is far to the left from
the points under the investigation and consists only of the waves moving in the
positive direction, then Eq.\ (\ref{tuntimeasympt}) may be simplified.

In the energy representation
\[
\hat{T}\left( x\right) =\int_{-\infty }^{\infty }t_{1}dt_{1}\sum_{\alpha
,\alpha ^{\prime }}\int\!\!\!\int dE\,dE^{\prime }\left| E,\alpha \right\rangle
\left\langle E,\alpha \right| \hat{J}\left( x\right) \left| E^{\prime
},\alpha ^{\prime }\right\rangle \left\langle E^{\prime },\alpha ^{\prime
}\right| \exp \left( \frac{i}{\hbar }\left( E-E^{\prime }\right)
t_{1}\right) .
\]
Integral over time is equal to $2i\pi \hbar ^{2}\frac{\partial }{\partial
E^{\prime }}\delta \left( E-E^{\prime }\right) $ and we obtain
\begin{eqnarray*}
\hat{T}\left( x\right) &=&-i\hbar 2\pi \hbar \sum_{\alpha ,\alpha ^{\prime
}}\int dE\left| E,\alpha \right\rangle \left( \left. \frac{\partial }{
\partial E^{\ \prime }}\left\langle E,\alpha \right| \hat{J}\left( x\right)
\left| E^{\prime },\alpha ^{\prime }\right\rangle \right| _{E^{\prime
}=E} \left\langle E,\alpha ^{\prime }\right| \right. \\
&& + \left. \left\langle E,\alpha \right| \hat{J}\left( x\right) \left| E,
\alpha ^{\prime}\right\rangle \frac{\partial }{\partial E}\left\langle E,
\alpha ^{\prime}\right| \right) , \\
\left\langle \hat{N}\left( X\right) \hat{T}\left( x\right) \right\rangle
&=&-i\hbar 4\pi ^{2}\hbar ^{2}\sum_{\alpha }\int dE\langle \Psi \left|
E,+\right\rangle \left\langle E,+\right| \hat{J}\left( X\right) \left|
E,\alpha \right\rangle \\
&& \times \left( \left. \frac{\partial }{\partial E^{\ \prime }}
\left\langle E,\alpha \right| \hat{J}\left( x\right) \left| E^{\prime
},+\right\rangle \right| _{E^{\prime }=E} + \left\langle E,\alpha \right|
\hat{J}\left( x\right) \left| E,+\right\rangle \frac{\partial }{\partial E}
\right) \left\langle E,+\right| \Psi \rangle .
\end{eqnarray*}
Substituting expressions for the matrix elements of the probability flux
operator we obtain equation
\begin{eqnarray*}
\left\langle \hat{N}\left( X\right) \hat{T}\left( x\right) \right\rangle
&=& \int dE\langle \Psi \left| E,+\right\rangle t^{*}\left( E\right) \frac{
\hbar }{i}\frac{\partial }{\partial E}t\left( E\right) \left\langle
E,+\right| \Psi \rangle \\
&& + Mx\int dE\langle \Psi \left| E,+\right\rangle \frac{%
1}{p_{E}}\left| t\left( E\right) \right| ^{2}\left\langle E,+\right| \Psi
\rangle \\
&& + i\hbar \frac{M}{2}\int dE\langle \Psi \left| E,+\right\rangle \frac{
1}{p_{E}^{2}}r^{*}\left( E\right) t^{2}\left( E\right) \exp \left( 2\frac{i}{
\hbar }p_{E}x\right) \left\langle E,+\right| \Psi \rangle .
\end{eqnarray*}

When $x\rightarrow +\infty $, the last term vanishes and we have
\begin{eqnarray}
\left\langle \hat{N}\left( X\right) \hat{T}\left( x\right) \right\rangle
& =& \int dE\langle \Psi \left| E,+\right\rangle t^{*}\left( E\right)
\frac{\hbar }{i}\frac{\partial }{\partial E}t\left( E\right) \left\langle
E,+\right| \Psi \rangle \nonumber \\
&& + Mx\int dE\langle \Psi \left| E,+\right\rangle \frac{%
1}{p_{E}}\left| t\left( E\right) \right| ^{2}\left\langle E,+\right| \Psi
\rangle,\quad x\rightarrow +\infty . \label{positivinfnty}
\end{eqnarray}
This expression is equal to $\left\langle \hat{T}\left( x\right)
\right\rangle $:
\begin{equation}
\left\langle\hat{N}\left(X\right)\hat{T}\left(x\right)
\right\rangle\rightarrow\left\langle\hat{T}\left(x\right)\right\rangle,
\quad x\rightarrow +\infty . \label{eq:lim}
\end{equation}

When the point with coordinate $x$ is in front of the barrier, we obtain
equality
\begin{eqnarray*}
\left\langle \hat{N}\left( X\right) \hat{T}\left( x\right) \right\rangle
&=& -i\hbar \int dE\langle \Psi \left| E,+\right\rangle \left| t\left(
E\right) \right| ^{2}\left( \frac{i}{\hbar }\frac{M}{p_{E}}x \right. \\
&& - \left. \frac{M}{2p_{E}^{2}}r\left( E\right) \exp \left( -\frac{i}{\hbar}
2p_{E}x\right) + \frac{\partial}{\partial E}\right) \left\langle E,+\right|
\Psi \rangle
\end{eqnarray*}
When $\left| x\right| $ is large the second term vanishes and we have
\begin{eqnarray}
\left\langle \hat{N}\left( X\right) \hat{T}\left( x\right) \right\rangle
& \rightarrow & Mx\int dE\langle \Psi \left| E,+\right\rangle \frac{1}{p_{E}}
\left| t\left( E\right) \right| ^{2}\left\langle E,+\right|
\Psi \rangle \nonumber \\
&& + \int dE\langle\Psi\left|E,+\right\rangle\left|t\left(E\right)\right|
^{2}\frac{\hbar }{i}\frac{\partial }{\partial E}\left\langle E,+\right| \Psi
\rangle . \label{negativinfnty}
\end{eqnarray}
The imaginary part of expression\ (\ref{negativinfnty}) is not zero. This
means, that for determination of the asymptotic time it is insufficient to
integrate only in the region containing the barrier. For quasimonochromatic
wave packets from Eqs.\ (\ref{opT}),\ (\ref{tuntimeasympt}),\
(\ref{asympttimeterm}), \ (\ref{positivinfnty}) and\ (\ref{negativinfnty}) we
obtain limits
\begin{mathletters}
\label{asymptmonochrom}
\begin{eqnarray}
t^{\text{Tun}}\left( x_{2},x_{1}\right)  &\rightarrow &t_{T}^{\text{Ph}}+
\frac{1}{p_{E}}M\left( x_{2}-x_{1}\right) , \\
t_{\text{corr}}^{\text{Tun}}\left( x_{2},x_{1}\right)  &\rightarrow &-
t_{T}^{\text{Im}}
\end{eqnarray}
\end{mathletters}
where
\begin{equation}
t_{T}^{\text{Ph}}=\hbar\frac{d}{dE}\left(\arg t\left(E\right)\right)
\label{phasetime}
\end{equation}
is the phase time and
\begin{equation}
t_{T}^{\text{Im}}=\hbar\frac{d}{dE}\left(\ln\left|t\left(E\right)
\right|\right)   \label{imagtime}
\end{equation}
is the imaginary part of the complex time.

In order to take the limit $x\rightarrow -\infty$ we have to perform more
exact calculations. We cannot extend the range of the integration over the time
to $-\infty$, because this extension corresponds to the initial wave packet being
infinitely far from the barrier. We can extend the range of the integration
over the time to $-\infty$ only for calculation of $\hat{N}\left(X\right)$. For
$x<0$ we obtain the following equality
\begin{equation}
\left\langle \hat{N}\left( X\right) \hat{T}\left( x\right) \right\rangle =
\frac{1}{4\pi Mi}\int_{0}^{\infty }tdt\left( I_{1}^{*}\left( x,t\right)
\frac{\partial}{\partial x}I_{2}\left( x,t\right) -I_{2}\left( x,t\right)
\frac{\partial}{\partial x}I_{1}^{*}\left( x,t\right) \right)
\end{equation}
where
\begin{eqnarray}
I_{1}\left( x,t\right) &=&\int dE\frac{1}{\sqrt{p_{E}}}\left| t\left(
E\right) \right| ^{2}\exp \left( \frac{i}{\hbar }\left( p_{E}x-Et \right)
\right) \left\langle E,+\right| \Psi \rangle , \\
I_{2}\left( x,t\right) &=&\int dE\frac{1}{\sqrt{p_{E}}}\left( \exp \left(
\frac{i}{\hbar }p_{E}x\right) +r\left( E\right) \exp \left( -\frac{i}{\hbar }
p_{E}x\right) \right) \exp \left( -\frac{i}{\hbar }Et\right)
\left\langle E,+\right| \Psi \rangle .
\end{eqnarray}
\begin{figure}[tbp]
\begin{center}
\epsfxsize=.5\hsize \epsffile{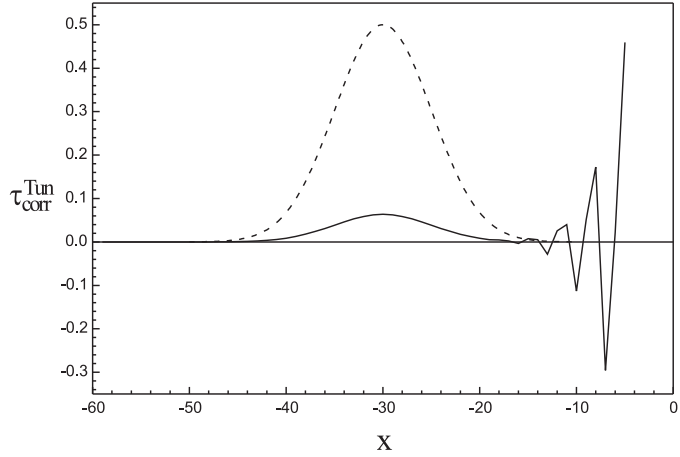}
\end{center}
\caption{The quantity $\tau_{\text{corr}}^{\text{Tun}}\left( x\right) $ for
$\delta $ function barrier with the parameters and initial conditions as in
Fig.\ \protect\ref{tundelta}. The initial packet is shown with dashed line.}
\label{deltabarjkomutat}
\end{figure}
\begin{figure}[tbp]
\begin{center}
\epsfxsize=.5\hsize \epsffile{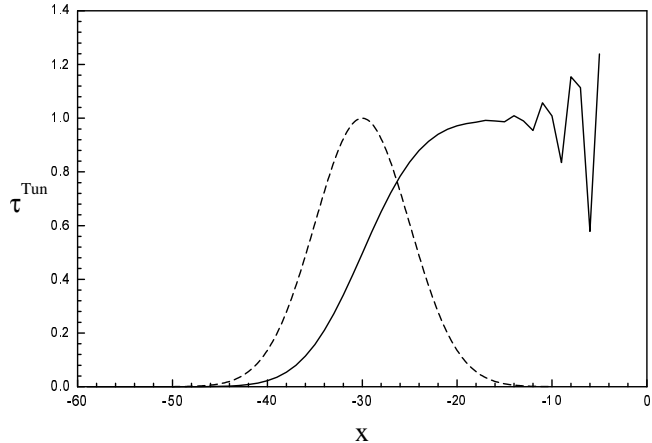}
\end{center}
\caption{``Tunneling time density'' for the same conditions and parameters as
in Fig.\ \protect\ref{deltabarjkomutat}.}
\label{deltabarjtikslus}
\end{figure}
$I_{1}\left(x,t\right)$ is equal to the wave function in the point $x$ at the
time moment $t$ when the propagation is in the free space and the initial wave
function in the energy representation is $\left|t\left(
E\right)\right|^{2}\left\langle E,+\right|\Psi\rangle$. When $t\geq 0$ and
$x\rightarrow -\infty$, then $I_{1}\left(x,t\right)\rightarrow 0$. That is why
the initial wave packet contains only the waves moving in the positive
direction. Therefore $\left\langle\hat{N}\left(X\right)\hat{T}\left(
x\right)\right\rangle\rightarrow 0$ when $x\rightarrow -\infty$. From this
analysis it follows that the region in which the asymptotic time is determined
has to contain not only the barrier but also the initial wave packet.

In such a case from Eqs.\ (\ref{tuntimeasympt}) and\ (\ref{asympttimeterm}) we
obtain expression for the asymptotic time
\begin{equation}
t^{Tun}\left(x_{2},x_{1}\rightarrow -\infty\right) =\frac{1}{\left\langle
\hat{N}\left( X\right) \right\rangle }\int dE\langle \Psi \left|
E,+\right\rangle t^{*}\left( E\right) \left( \frac{M}{p_{E}}x_{2}-i\hbar
\frac{\partial }{\partial E}\right) t\left( E\right) \left\langle E,+\right|
\Psi \rangle . \label{mainresult}
\end{equation}
From Eq. (\ref{eq:lim}) it follows that
\begin{equation}
t^{Tun}\left(x_{2},x_{1}\rightarrow -\infty\right)=\frac{1}{\left\langle
\hat{N}\left(X\right) \right\rangle}\left\langle\hat{T}
\left(x_{2}\right)\right\rangle \label{eq:fin}
\end{equation}
where $\hat{T}\left(x_{2}\right)$ is defined as the probability flux
integral\ (\ref{opT}). Eqs.\ (\ref{mainresult}) and\ (\ref{eq:fin}) give
the same value for tunneling time as an
approach in Refs.\cite{delgadomuga,grotrovelli}

The integral of quantity $\tau_{\text{corr}}^{\text{Tun}}\left(x\right)$ over
a large region is zero. We have seen that it is not enough to choose the
region around the barrier---this region has to include also the initial wave
packet location. We illustrate this fact by numerical calculations.

The quantity $\tau_{\text{corr}}^{\text{Tun}}\left(x\right)$ for
$\delta$-function barrier is represented in Fig. \ref{deltabarjkomutat}. We see
that $\tau_{\text{corr}}^{\text{Tun}}\left(x\right)$ is not equal to zero not
only in the region around the barrier but also it is not zero in the location
of the initial wave packet. For comparison, the quantity
$\tau^{\text{Tun}}\left(x\right)$ for the same conditions is represented in
Fig. \ref{deltabarjtikslus}.

\section{Conclusion}
\label{secconcl}

We have shown that it is impossible to determine the time a tunneling particle
spends under the barrier, because the knowledge about the location of the
particle is incompatible with the knowledge whether the particle will tunnel or
not. This is because the corresponding operators, given by Eqs.\
(\ref{tunflagheis}) and\ (\ref{posopheis}) do not commute. However, it is
possible to speak about the asymptotic time, i.e., the time the particle spends
in a large region.

In order to illustrate those facts, to obtain an expression of the asymptotic
time and to investigate its behavior, we consider a procedure of time
measurement, proposed by A. M. Steinberg \cite{steinberg}. This procedure shows
clearly the consequences of non-commutativity of the operators and the
possibility of determination of the asymptotic time. Our model also reveals the
Hartmann and Fletcher effect, i.e., for opaque barriers the effective velocity
is very large, because the contribution of the barrier region to the time is
almost zero. We cannot determine whether this velocity can be larger than $c$,
because for this purpose one has to use a relativistic equation (e.g., the
Dirac's equation).

Due to non-commutativity of operators\ (\ref{tunflagheis})
and\ (\ref{posopheis}) the outcome of measurements depends on particular
detector even in an ideal case. This makes the measurement of the tunneling
time difficult for opaque barriers, because the tunneling time is very short
and the term depending on the detector increases linearly with the barrier
width. This term vanishes when the time spent in a large region, including
initial packet location, is measured.

\acknowledgments

I wish to thank Prof. B. Kaulakys for his suggestion of the problem, for
encouragement, stimulating discussions and critical remarks. I am also indebted
to the referee for the useful comments and suggestions for the improvement of
this work.

\end{document}